\documentclass[a4paper,11pt]{article}
\usepackage{pos}
\usepackage{subcaption}
\usepackage{hyperref}

\hypersetup{
    colorlinks=true,
    linkcolor=blue,
}

\title{A numerical and theoretical study of multilevel performance for two-point correlator calculations}
\ShortTitle{A study of multilevel performance for two-point correlators}

\author*[a]{Ben Kitching-Morley}
\author[a,b]{Andreas J\"uttner}

\affiliation[a]{Mathematical Sciences, University of Southampton, Highfield,\\
Southampton SO17 1BJ, United Kingdom}
\affiliation[a, b]{School of Physics and Astronomy, University of Southampton,\\
Southampton SO17 1BJ, United Kingdom}
\affiliation[a, b]{STAG Research Center, University of Southampton, Highfield,\\ Southampton SO17 1BJ, United Kingdom}

\affiliation[b]{Theoretical Physics Department, CERN, 1211 Geneva 23, Switzerland}

\emailAdd{bkm1n18@soton.ac.uk}

\abstract{An investigation of the performance of the multilevel algorithm in the approach to criticality has been undertaken using the Ising model, performing simulations across a range of temperatures. Numerical results show that the performance of multilevel in this system deteriorates as the correlation length is increased with respect to the lattice size. The statistical error of the longest correlator in the system is reduced in a multilevel setup when the correlation length is less than one-tenth of the lattice size, while for longer correlation lengths multilevel performs more poorly than a computer-time equivalent single level algorithm. A theoretical model of this performance scaling is outlined, and shows remarkable accuracy when compared to numerical results. This theoretical model may be applied to other systems with more complex spectra to predict if multilevel techniques are likely to result in improved statistics.}

\FullConference{%
 The 38th International Symposium on Lattice Field Theory, LATTICE2021
  26th-30th July, 2021
  Zoom/Gather@Massachusetts Institute of Technology
}

\begin{document}
\maketitle

\section{Introduction}
\noindent One barrier to precise lattice results is the signal-to-noise problem. It occurs when the two-point correlator decays exponentially with increasing separation while the statistical error due to random fluctuations remains constant. Multilevel techniques (\cite{Parisi:1983hm}, \cite{Luscher:2001up}) involve dividing the lattice into sub-regions separated by boundaries and simulating these sub-regions independently. They have proven effective at overcoming the signal-to-noise problem, especially in lattice gauge theory. There is a great deal of interest in studying theories at criticality. Of particular interest to the authors are holographic models of the early universe \cite{McFadden:2009fg}. To simulate these theories one must tune a bare mass parameter such that the theory is at a nonperturbative massless (critical) point \cite{Cossu:2020yeg}. In this study numerical results from the Ising model are used to study the performance of multilevel in the approach to criticality. Additionally, a theoretical model to understand this performance is proposed and tested against this numerical data.

\section{The multilevel algorithm}
\noindent In this article we will focus on a two-level setup, splitting the lattice $\Lambda$ into two sub-regions $\{\Lambda_1, \Lambda_2\}$ which are separated by boundaries $\partial B$. The action is local so with boundaries one lattice site thick there are no contributions that mix $\Lambda_1$ and $\Lambda_2$. The path integral can then be decomposed as
\begin{align}
   \int_{x \in \Lambda} \mathcal{D}\phi(x) e^{-S[\Lambda]} = \textcolor{red}{\int_{x \in \partial B} \mathcal{D}\phi(x) e^{-S[\partial B]}} \prod_{r = 1}^2 \int_{x \in \Lambda_r} \mathcal{D}\phi(x) e^{-S[\Lambda_r; \textcolor{red}{\partial B}]}. \label{eq: factorization of path integral}
\end{align}
In performing a multilevel simulation we first produce $N$ configurations of the whole lattice. These configurations are used to fix the boundary sites $\partial B_i$ where $i \in \{1, 2, ..., N\}$. We produce sub-lattice ensembles, with $M$ configurations for each boundary configuration, labelled by the index $j_r \in \{1, 2, ..., M \}$ for sub-region $\Lambda_r$. In the following calculations we assume that the Monte-Carlo time between successive boundary configurations and between successive sub-lattice configurations is sufficiently large that the effects of autocorrelation can be ignored. Consider fields insertions $\phi(x)$ and $\phi(y)$, where $x$ is in sub-lattice $\Lambda_1$ and $y$ is in $\Lambda_2$. For a given boundary configuration $\partial B_i$ the field values $\phi(x)_{ij_1}$ and $\phi(y)_{ij_2}$ are sampled from distributions with means $\mu_x(\partial B_i)$ and $\mu_y(\partial B_i)$ and variances $\sigma^2_x(\partial B_i)$ and $\sigma^2_y(\partial B_i)$ respectively. Defining
\begin{align}
    X_i = \frac{1}{M}\sum_{j_1} \phi(x)_{ij_1},\qquad Y_i = \frac{1}{M}\sum_{j_2} \phi(y)_{ij_2},
\end{align}
we use the central limit theorem to give us
\begin{align}
    X_i \sim N\left(\mu_x(\partial B_i), \frac{\sigma^2_x(\partial B_i)}{M}\right), \qquad Y_i \sim N\left(\mu_y(\partial B_i), \frac{\sigma^2_y(\partial B_i)}{M}\right).
\end{align}
The two-point correlator is given by $Z = (1/N)\sum_i \tilde{Z}_i / N$, where $\tilde{Z}_i = X_i Y_i$. The two-point correlation between $\phi(x)$ and $\phi(y)$ is accounted for through the means of their distributions ($\mu_x(\partial B_i)$ and $\mu_y(\partial B_i)$), which are both conditionally dependent on the boundary. There is no residual statistical correlation between $X_i$ and $Y_i$. We therefore have that
\begin{align}
    \langle \tilde{Z}_i \rangle &= \langle X_i Y_i \rangle =\langle X_i \rangle \langle Y_i \rangle = \mu_{X_i} \mu_{Y_i},
\end{align}
where $\mu_{X_i} = \mu_x(\partial B_i)$ and $\mu_{Y_i} = \mu_y(\partial B_i)$. One can show that the variance of $\tilde{Z}_i$ is given by
\begin{align}
\sigma^2_{\tilde{Z}_i} = \sigma^2_{X_i} \sigma^2_{Y_i} + \mu_{X_i}^2 \sigma^2_{Y_i} + \mu_{Y_i}^2 \sigma^2_{X_i},
\end{align}
where $\sigma^2_{X_i} = \sigma^2_x(\partial B_i)/M$ and $\sigma^2_{Y_i} = \sigma^2_y(\partial B_i)/M$. Using $E$ and $Var$ to represent the expectation value and variance of $\tilde{Z}$ as we vary the boundary configuration, the law of total variance tells us that
\begin{align}
    Var(\tilde{Z}) &= Var(\mu_{\tilde{Z}}) + E(\sigma^2_{\tilde{Z}}), \\
    &= Var(\mu_X \mu_Y) + E(\mu^2_X \sigma^2_Y + \mu^2_Y \sigma^2_X + \sigma^2_X \sigma^2_Y), \nonumber
\end{align}
giving us overall that (see also \cite{GarciaVera:2017xif}))
\begin{align}
    Var(Z) &= \frac{1}{N}\left( Var(\mu_x \mu_y) + \frac{1}{M}E(\mu^2_x \sigma^2_y + \mu^2_y \sigma^2_x) + \frac{1}{M^2}E(\sigma^2_x \sigma^2_y)\right). \label{eq: Variance General}
\end{align}
The relative contribution of the terms in this formula will determine if a multilevel algorithm outperforms a single level one. The computational cost of a multilevel simulation is equivalent to a single level simulation with $N \times M$ configurations, which has a variance scaling like $1 / NM$. If the two-point function is largely boundary dominated, then $Var(\mu_x \mu_y) / N$ will be the leading scaling, and a multilevel algorithm will give correlator estimates with a variance $M$ larger than an equivalent single level algorithm. By contrast, if the boundary has little impact on $\phi(x)$ and $\phi(y)$, and these field insertions have zero mean, we achieve a $1/NM^2$ scaling of the variance, which is an improvement over single level variance scaling by a factor of $M$.\\

Multilevel techniques are now used widely in the simulation of lattice gauge theories, however their application in other areas of lattice physics is an area of active study. Research has been done to implement multilevel in QCD with fermions where the quark propagators are non-local and thicker boundaries must be used \cite{Ce:2016ajy}. Multilevel algorithms has also been studied as a potential technique to overcome critical slowing down \cite{Jansen:2020qrz}. Some authors have investigated the use of symmetries to improve the efficiency of multilevel \cite{DellaMorte:2010yp}. Recently multilevel methods have been used in calculations of the muon magnetic moment \cite{DallaBrida:2020cik}. \\

\begin{figure}
    \centering
    \includegraphics[width=80mm]{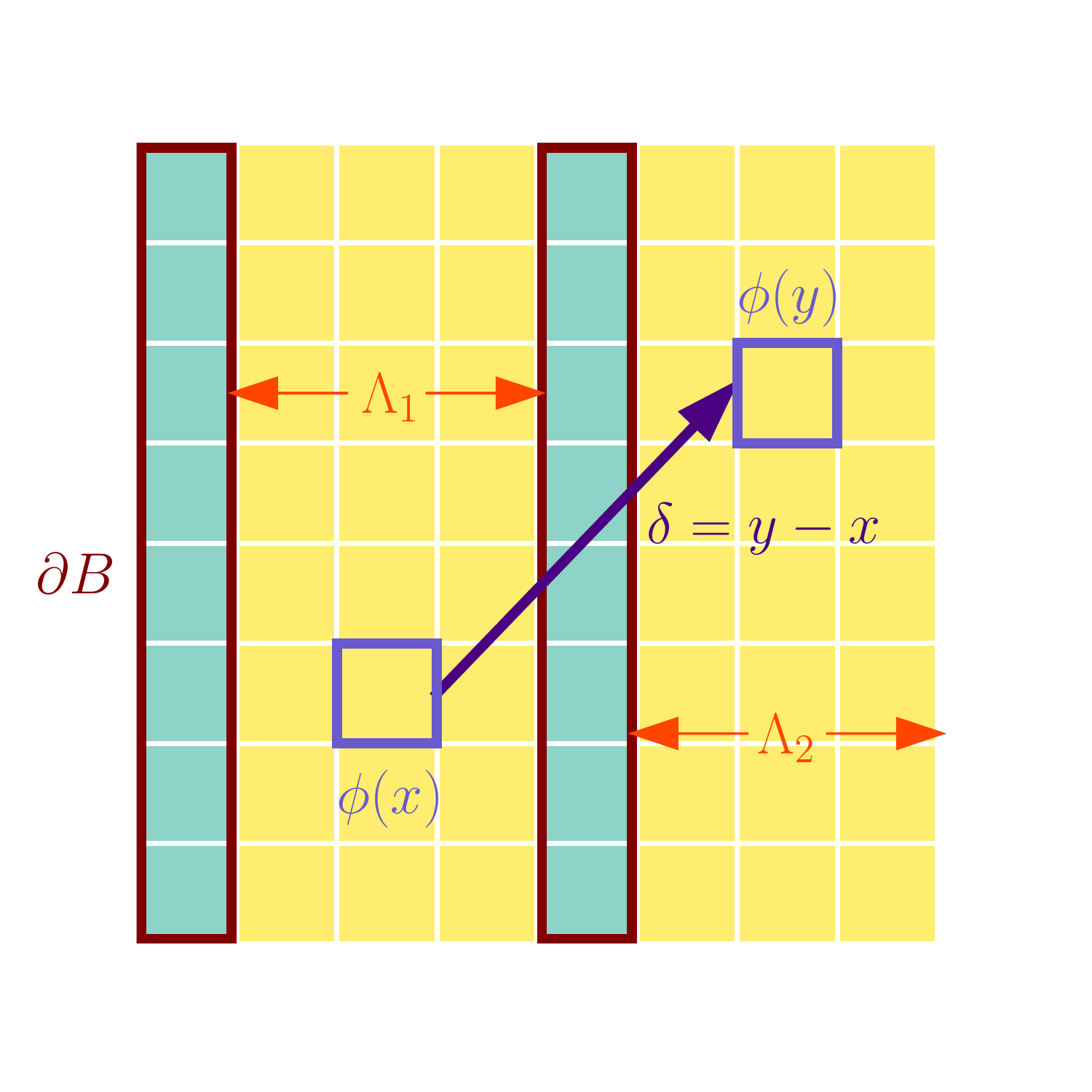}
    \caption{A multilevel set up with two sub-lattices.}
    \label{fig: multilevel diagram}
\end{figure}

In this paper the two-dimensional Ising model has been used as a test system to investigate the performance of multilevel in the approach to criticality. Fields $\phi((x, y))$ populate a two-dimensional square lattice of size $L$, and have values in the set $\{-1, +1\}$. The discretized path integral is
\begin{align}
    Z = \int \mathcal{D}\phi \, exp\left( -\beta \left( J \sum_{(i, j) n.n.}\phi_i \phi_j + B \sum_i \phi_i\right)\right), \label{eq: Ising path integral}
\end{align}
where the first term involves a sum over nearest neighbours (n.n.) and the second term is the energy due to an external magnetic field, $B$. We take $J = 1$, $B = 0$ from here on. In this instance the system has a known second-order phase transition, with a critical temperature $T_c = \left(\frac{1}{2}\log(1 + \sqrt{2})\right)^{-1}$ \cite{Onsager:1943jn}. The theory was simulated using a Metropolis-Hastings \cite{Metropolis:1953am, Hastings:1970aa} algorithm implemented in Python. The performance of the simulation could be improved by the use of a compiled language, parallelization and clustering techniques \cite{Wolff:1988uh}. However, the Ising model is being used here only as a test system to investigate multilevel and hence the precision of the results isn't intended to be competitive. For simplicity we use slice-coordinate fields from now on, $\Phi(x) = \frac{1}{L}\sum_y \phi((x, y))$. We split the lattice along the x-axis so that a given slice-coordinate field only contains contributions from $\partial B$, $\Lambda_1$ or $\Lambda_2$.

\section{Evaluating multilevel performance}
\noindent To evaluate the performance of our multilevel simulations we compare them to a computationally equivalent single level simulation with $N_{\rm{single}} = N \times M$ configurations. In this system the two-point correlator of slice coordinates at $x$ and $x + \delta$ is given by
\begin{align}
    C^s(\delta; x) &= \frac{1}{N_{\rm{single}}} \sum_{i = 1}^{N_{\rm{single}}} \Phi_i(x) \Phi_i(x + \delta).
\end{align}
If $x + \delta \geq L$ we apply periodic boundary conditions: $\Phi_i(x + \delta) = \Phi_i(x + \delta - L)$. The multilevel two-point function between a slice-coordinate field at $x \in \Lambda_r$ and $x + \delta \in \Lambda_s$ is given by
\begin{align}
    C^m(\delta; x) &= \frac{1}{NM^2} \sum_{i=1}^{N} \sum_{j_r = 1}^M \sum_{j_s = 1}^M \Phi_{ij_r}(x) \Phi_{ij_s}(x+ \delta),
\end{align}
where $i$ indexes the boundary configuration and $j_k$ indexes the sub-lattice configuration in region $\Lambda_k$. Note that because the multilevel decomposition of the path integral is exact, $E(C^m(\delta; x)) = E(C^s(\delta; x)) := C_2(\delta)$. After performing an average over the boundary configurations, we can apply the central limit theorem,
\begin{align}
    C_2^s(\delta; x) \sim N\left(C_2(\delta), \sigma_s^2(x)\right), \qquad C_2^m(\delta; x) \sim N\left(C_2(\delta), \sigma^2_m(x)\right),
\end{align}
where $\sigma^2_m(x)$, in the case of $\Lambda_r \neq \Lambda_s$, is given by equation (\ref{eq: Variance General}) and in other cases by similar expressions. Analogous expressions hold for the single level variance $\sigma^2_s(x)$.

\subsection{Optimum Weighting}
\noindent To obtain the overall estimate for the two-point correlator we take a weighted average across all values of $x$,
\begin{align}
    C^s_2(\delta) &= \sum_x W^s_x C^s_2(\delta; x),\\
    C^m_2(\delta) &= \sum_x W^m_x C^m_2(\delta; x), \nonumber
\end{align}
where $\sum_x W^s_x = \sum_x W^m_x =1$. Defining $\textbf{W}^m = (W^m_1, W^m_2, ..., W^m_L)$ and $\textbf{W}^s = (W^s_1, W^s_2, ..., W^s_L)$, the overall distribution of our correlator estimator is given by
\begin{align}
    C_2^s(\delta) &\sim \left(C_2(\delta), \textbf{W}^s \cdot Cov^s(\delta) \cdot \textbf{W}^s\right),\\
    C_2^m(\delta) &\sim \left(C_2(\delta), \textbf{W}^m \cdot Cov^m(\delta) \cdot \textbf{W}^m\right),
\end{align}
where $Cov^m(\delta)$ is the $L \times L$ covariance matrix between multilevel correlators of separation $\delta$ with their first insertion at different positions on the lattice: $Cov^m(\delta)_{x_1x_2} = \langle C_2^m(\delta; x_1) C_2^m(\delta; x_2) \rangle - \langle C_2^m(\delta; x_1) \rangle \langle C_2^m(\delta; x_2) \rangle $, while $Cov^s(\delta)$ is defined similarly. We choose the values of $W^m_x$ that minimize the quadratic form $\textbf{W}^m \cdot Cov^m(\delta) \cdot \textbf{W}^m$ subject to $\sum_x W^m_x = 1$. For a single level algorithm the optimal choice of $\mathbf{W}^s$ is $W^s_x = 1/L \: \forall \: x$. In this piece of work, the covariance matrix has been determined in two different ways. The first is to use the simulation data to calculate an empirical covariance matrix, which is used to weight the correlators and numerically evaluate multilevel. To avoid introducing a bias, the data used to find optimal weights was separated from the data weighted by those weights. We therefore split the data into sets, and use the data in all but one of the sets to determine the weighting for the data in the remaining set. We repeat this for each set in turn to get weighted contributions from all sets. The second part of this work is to provide a model for predicting the covariance matrix, and therefore multilevel performance.

\subsection{Observed Performance Gain}
\begin{figure}
    \centering
    \includegraphics[width=140mm]{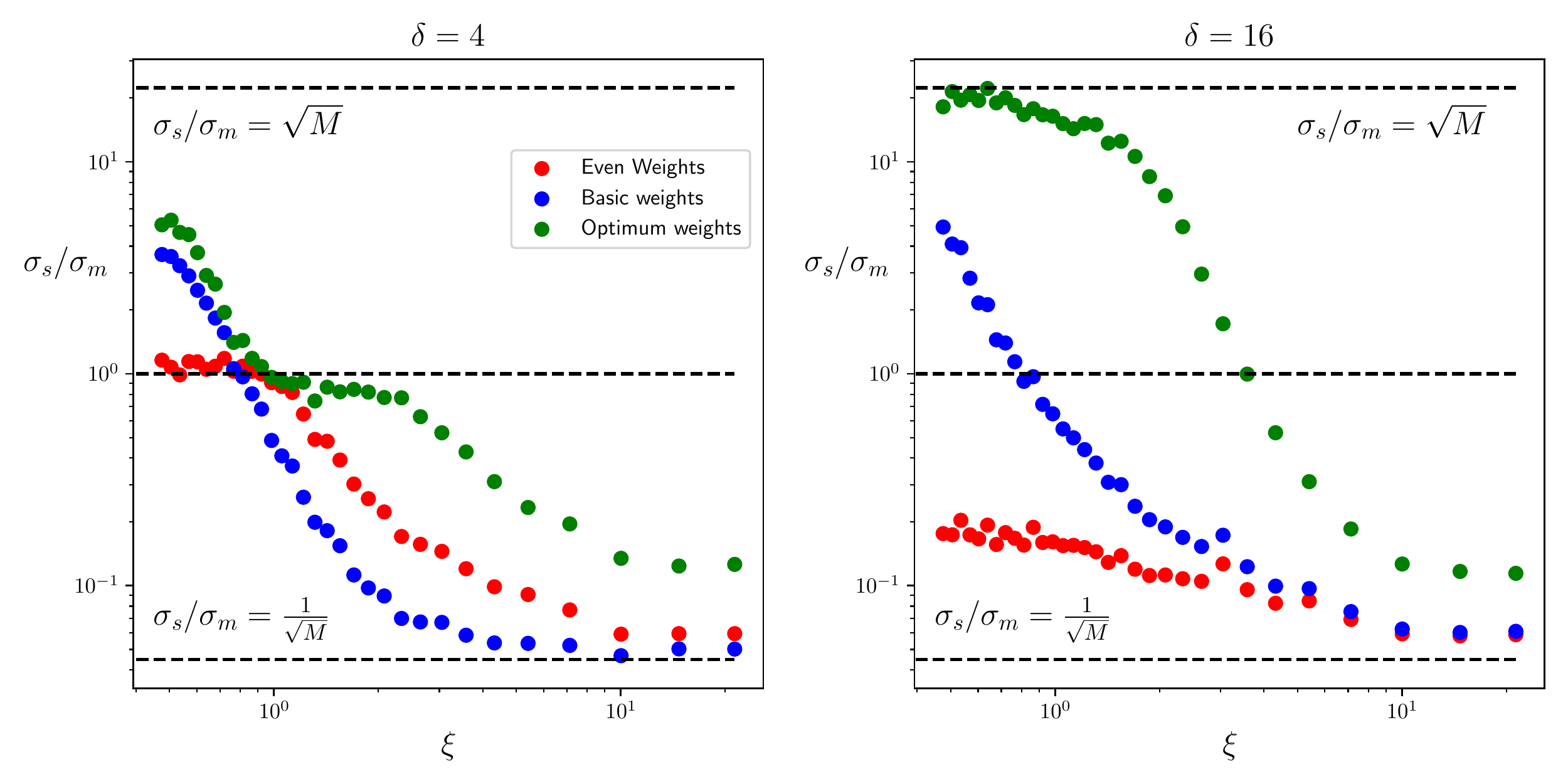}
    \caption{The ratio of the standard deviation of a single level simulation and a multilevel simulation of equivalent computational time is shown against the correlation length. Here $N = 500$ boundary configurations and $M = 500$ sub-lattice configurations were used on a $32 \times 32$ lattice. The black-dashed lines represent the theoretical best and worst performance scenarios for multilevel.}
    \label{fig: Multilevel performance diff delta}
\end{figure}

\noindent An $L=32$, $N=500$, $M=500$ multilevel simulation and a computational-cost-equivalent single level setup were both executed using Python. The ratio $\sigma_{s}/\sigma_{m}$ against the correlation length $\xi$ is shown for both short ($\delta = 4$) and long ($\delta = 16$) correlators in figure (\ref{fig: Multilevel performance diff delta}). For the single level system a uniform weighting was used, while for the multilevel system one of three possible weighting schemes was used:

\begin{enumerate}
    \item \textbf{Optimum Weights} Weights that minimize the quadratic form $\textbf{W}^m \cdot Cov^m(\delta) \cdot \textbf{W}^m$ subject to $\sum_x W_x^m = 1$.
    \item \textbf{Even Weights} $W^m_x = 1/L, \: \forall \: x$.
    \item \textbf{Basic Weighting} Correlators between two boundaries have a weight of $1$, while correlators between a boundary and a non-boundary site, or between two sites in the same sub-lattice, have a weight of $M$. Correlators between two different sub-lattices have a weight of $M^2$. These weights are then normalized by $\sum_x W^m_x = 1$.
\end{enumerate}

When the correlation length is small compared to the lattice size a multilevel algorithm performs better than an equivalent single level one, while for larger correlation lengths it performs more poorly. When the correlation length is large, the sub-lattice field values are almost entirely determined by the boundary meaning the $1 / N$ scaling term in eq. (\ref{eq: Variance General}) dominates, causing the standard deviation of correlators to be $\sqrt{M}$ larger compared to the single level algorithm. With a small correlation length, the sub-lattice sites are largely independent of the boundary. Since we have $N \times M^2$ such contributions to the multilevel average, compared to the $N \times M$ for the single level algorithm, the gives up to a $\sqrt{M}$ reduction to the standard deviation. This theoretical upper bound in performance is approached by the $\delta = 16$ correlator, but not by the $\delta = 4$ correlator, as the shorter correlator involves multilevel interactions between field insertions closer to the boundary, and fewer contributions in general. In the limit $\xi \xrightarrow[]{} 0$ the second effect is dominant, and multilevel performance is $\sqrt{2(\delta - 1) /L}$ poorer than expected. This behavior is acceptable as the longest correlators suffer most from the signal-to-noise problem.\\

\begin{figure}
    \centering
    \includegraphics[width=120mm]{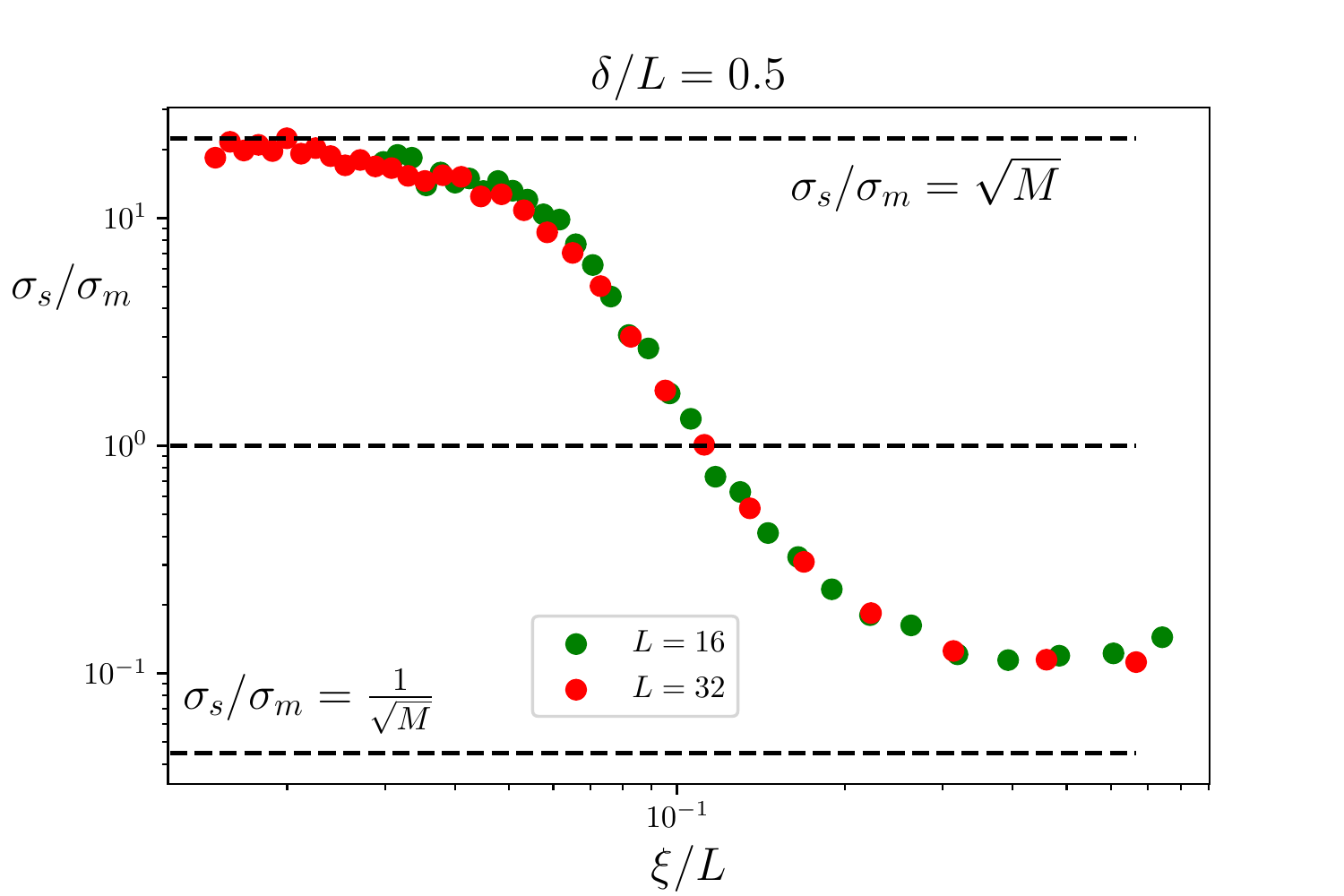}
    \caption{Variation of the performance of a two-level multilevel algorithm for $L = 16$ and $L = 32$ lattices with $r_1 = \xi / L$. Using a $\delta = L / 2$ correlator with $N = 500$ boundary configurations and $M = 500$ sub-lattice configurations.}
    \label{fig: scaling multilevel}
\end{figure}

We hypothesize that multilevel performance depends only on the ratios $r_1 = \xi / L$, and $r_2 = \delta / L$, or alternatively other pairs of unitless ratios formed by $\xi$, $\delta$ and $L$. This hypothesis was tested by comparing the multilevel performance gain for two different lattice sizes, where the ratio $r_2$ is kept constant, iterating over different $r_1$ values (fig. (\ref{fig: scaling multilevel})). Here, $L=16$ and $L=32$ systems are compared, with $r_2 = 0.5$. As expected the curves sit on top of eachother. In the regime where the size of the lattice $L$ is significantly larger than the correlation length, then the most relevant ratio is $\delta / \xi$.

\section{A theoretical model of multilevel performance}
\begin{figure}
    \centering
    \includegraphics{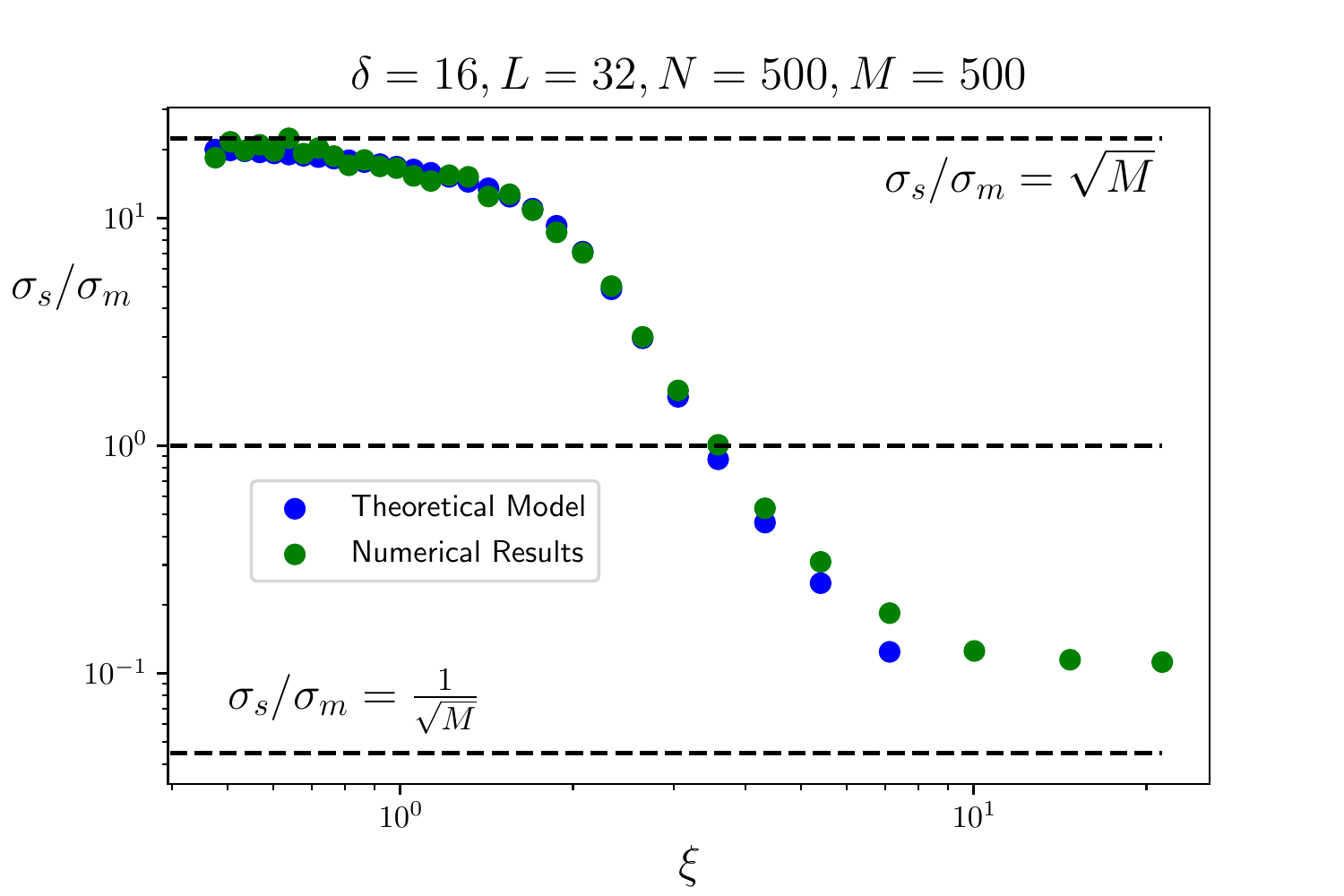}
    \caption{Performance gain of a two-level multilevel algorithm against correlation length as obtained by numerical observations, and as predicted by the theoretical model of performance. In the numerical simulation a lattice of size $32$ was used with a correlator separation of $16$, $500$ boundary configurations and $500$ sub-lattice configurations.}
    \label{fig: Theoretical Performance}
\end{figure}

\noindent To calculate theoretically the standard deviation of two-point correlators in our simulations, we must predict the covariance matrices $Cov^s(\delta)$ and $Cov^m(\delta)$. For example, in a single level setup, we consider four slice-coordinate field insertions at $x = S$, $x = S'$, $x = T$ and $x = T'$, labelling them by $\phi^R_i = \phi_i(x = R)$ where $i$ indexes the boundary configuration and $R \in \{1, 2, 3, ..., L\}$. We are interested in two-point correlators so we take $S' = S + \delta$ and $T' =  T + \delta$ giving,
\begin{align}
    C^{SS'}_2 = \frac{1}{N} \sum_{i = 1}^{N_{single}} \phi_i^S \phi_i^{S'}, \qquad 
    C^{TT'}_2 = \frac{1}{N} \sum_{i = 1}^{N_{single}} \phi_i^T \phi_i^{T'}.
\end{align}
The covariance between these two-point correlators is given by
\begin{align}
    Cov(C^{SS'}_2, C^{TT'}_2) &= \langle C^{SS'}_2 C^{TT'}_2 \rangle - \langle C^{SS'}_2 \rangle \langle C^{TT'}_2 \rangle = \frac{1}{N}\langle \phi^S \phi^{S'} \phi^T \phi^{T'} \rangle - C_2(\delta)^2.
\end{align}
In what follows it will be convenient to normalize the fields, $\varphi^R = \phi^R / \sigma_\phi$ so that $\sigma_\varphi = 1$. This normalization will however cancel when we take the ratio $\sigma_{single} / \sigma_{multi}$. We use the equation for the two-point correlator in the symmetric phase,
\begin{align}
    \alpha := \exp{\left(-\frac{|P -  Q|}{\xi}\right)} = C_2(|P - Q|).
\end{align}
We then enforce $\varphi^Q = \alpha \varphi^P + f(\alpha) \epsilon$, where $\epsilon \sim N(0, 1)$, so that $\langle \varphi^P\varphi^Q\rangle = C_2(|P - Q|)$. Taking $\varphi^P\sim N(0, 1)$ and $\varphi^Q\sim N(0, 1)$ we can show that $f(\alpha) = \sqrt{1 - \alpha^2}$. We perform this decomposition for $\varphi^S$ and $\varphi^T$ then repeat it when we add additional fields. For multilevel correlators it's necessary to add boundary fields into the system first, and correlate the fields in the sub-lattices to them. Further details of this model will be published in a later publication. This model performs extremely well in predicting the numerically observed multilevel performance (see figure (\ref{fig: Theoretical Performance})). As the correlation length is increased the slice-coordinate fields become less normally distributed and the assumptions of the model no longer hold.

\section{Conclusions and Outlook}
\noindent In this work the performance of the multilevel algorithm in calculating two-point functions has been investigated in the two-dimensional Ising model. As expected we observe that the performance of the algorithm is highly dependent on the correlation length, with a $\sqrt{M}$ improvement of statistics compared to a computationally-equivalent single level algorithm in the limit $\xi / L \xrightarrow[]{} 0$. As correlation length is increased this performance improvement is decreased, until there is a crossover regime around $\xi / L = 0.1$, above which multilevel performs more poorly than single level. A theoretical model of this algorithmic performance has been outlined, and provides an excellent description of the observed performance. This model doesn't directly use the action of the Ising model, but instead makes use of the functional form of the two-point function. This work could be extended by applying this model of multilevel performance to other systems with more complex spectra, for example in Lattice QCD, where the development of multilevel techniques is being actively researched.

\section{Acknowledgements}
\noindent A. J. acknowledges funding from STFC consolidated grants ST/ P000711/1 and ST/T000775/1. B. K. M. was supported by
the EPSRC Centre for Doctoral Training in Next Generation Computational Modelling Grant No.
EP/L015382/1.

\newpage

\end{document}